\documentclass[preprint,12pt]{elsarticle}

\usepackage{graphicx}
\usepackage{amssymb}
\usepackage{amsthm}
\usepackage{amsmath}
\usepackage{booktabs}
\usepackage{epstopdf}

\journal{Astroparticle Physics}

\begin{document}
\begin{frontmatter}

\title{Average Inhomogeneities in Milky Way SNII and The PAMELA Anomaly}

\author{James Stockton}

\address{Dept. of Astronomy, New Mexico State University, Las Cruces, NM 88001}

\ead{stockton@nmsu.edu} 

\begin{abstract}
A model is presented to estimate the fraction of Supernova Type-II events (SNII) occurring inside vs. outside a spiral arm for a given star formation episode.  The probability distribution function (PDF) for this fraction is given for use in models similar to those of Shaviv et al. \cite{Shaviv_2009}\cite{Piran_2009}. The calculated PDF for the SNII fraction, $SNII_{in/total}$, defined as the number of SNII inside a spiral arm divided by the total number of SNII from a star formation event, provides a constraint on the magnitude of supernova remnant (SNR) concentrations used in cosmic ray propagation models attempting to explain the PAMELA anomaly. Despite the concentration of star formation within spiral arms, this model predicts the majority of SNII events actually occur in inter-arm regions and calls into question the SNR concentration assumption of Shaviv et al.

\end{abstract}


\end{frontmatter}

\section{Introduction}
Recently, the PAMELA collaboration released their results for the Galactic cosmic ray (GCR) positron fraction, $J_{e^{+}}/(J_{e^{-}}+J_{e^{+}})$\cite{PAMELA_2009}.  The remarkable feature in the PAMELA results is an increase in the positron fraction at energies above $\sim10$ Gev. Standard GCR propagation models assume all positrons are secondarily produced via interactions with the interstellar medium (ISM). This results in a decreasing positron fraction with energy.

Many models have been proposed to account for this discrepancy \cite{PAMELA2_2009}: DM particle annihilations, reacceleration of secondaries in supernova remnants (SNRs) (eg. \cite{Ahlers_2009}), inhomogeneities in the distribution of SNRs (eg. \cite{Shaviv_2009}\cite{Piran_2009}). For a more thorough discussion of proposed models and their implications, see \cite{Fan_2010}. Models relying on a concentration of SNR in the spiral arms of the Milky Way (MW) relative to the inter-arm regions have reproduced many observed results \cite{Piran_2009}\cite{Shaviv_2009}. However, the literature is sparse on the specific value of this concentration; thus, it has been difficult to constrain the magnitude of the concentrations used to plausible ranges.

In section 3, I present the model used to calculate $SNII_{in/total}$ as well as the necessary input parameters. Section 4 presents the results of the calculations and the PDF of $SNII_{in/total}$. Section 5 provides discussion on the use of the results and possible refinements.

\section{Model and Inputs}
If we assume a dense cloud in the ISM collapses to form a cluster of stars upon its first contact with the density enhancement of a spiral arm we can compare the time needed for a newly formed star to orbit through and exit the arm to the lifetimes of stars in the cluster. The spread in stellar velocities and lifetimes naturally leads to some SNII events occurring within the spiral arm and some occurring after the progenitor has orbited out of the arm.

Calculating the ratio of SNII events inside a spiral arm to the total number of SNII expected from a single star formation event, $SNII_{in/total}$, is accomplished by finding the ratio of two areas in the initial mass function (IMF) for star formation. The two areas are defined by 1., the mass above which stars explode in SNII, and 2., the mass at which stars explode via SNII while still within the spiral arm.  This second mass-point on the IMF is a function of several parameters.

Given that the pattern speed of the spiral density wave is lower than the orbital speeds of stars near the Solar Galactocentric radius, recently formed stars will, with time, exit the spiral arm along its leading edge (See Figure 1). However, if a given star were massive enough it would evolve through its lifecycle and end in a SNII while still within the spiral arm. Thus, to calculate $SNII_{in/total}$ we need to determine the fraction of stars with masses high enough to eventually explode as SNII that are also massive enough to do so prior to existing to confines of the spiral arm.

Thus, calculating the SNII fraction requires knowledge of 3 observationally constrained parameters: 1) the shape of the IMF \cite{Kroupa_2005};  2) the mass at which stars end their lives via SNII \cite{Smartt_2009};  3) the mass at which stars go SNII within the spiral arm, which depends on Solar Galactocentric radius \cite{Reid_2009}, Spiral Pattern Speed \cite{Grosbol_2010},\cite{Dobbs_2010}, orbital speeds of newly formed stars \cite{Reid_2009}, the size of the spiral arm \cite{Hou_2009},\cite{Anderson_2009}, and stellar lifetime as a function of stellar mass \cite{Hurley_2000}.

Each of these input parameters has an associated error. In order to address this,  I evenly sampled the error range for each parameter and calculated $SNII_{in/total}$ for all possible combinations of sampled values. This maps out the changes in $SNII_{in/total}$ with respect to errors in all input parameters. The distribution of calculated $SNII_{in/total}$ values, when normalized to area=1.0, provides a probability distribution for $SNII_{in/total}$. 

   In addition to the ranges and iteration values given in Table 1, the conversion of stellar mass to stellar lifetime was performed for stellar metalicities of Z=[0.001],[0.02],[0.03] \cite{Hurley_2000}.

\begin{table}[htbp]
   \centering
   \begin{tabular}{@{} lccr @{}}
      \toprule
      \multicolumn{3}{c}{Input Summary} \\
      \cmidrule(){1-3} 
      Parameter & Range & Iterations\\
      \midrule
       SNII Mass Limit, $M_{\odot}$ & 7.0-9.5 & 11 \\
       High Mass IMF Slope &  2.3-2.7 & 9 \\
       Galoctcentric Solar Radius, kpc &  7.9-9.0 & 12 \\
       Solar Orbital Speed, km/s &  223.0-255.0 & 9  \\
       Spiral Pattern Speed, km/s/kpc &   5.0-18.0 & 14  \\
       Spiral Arm Size, kpc & 0.5-3.0 & 14  \\
                            	                     
      \bottomrule
   \end{tabular}
   \caption{Each input parameter was stepped evenly through its observational range to produce the listed number of iterations}
   \label{tab:booktabs}
\end{table}

\section{Results}

	\begin{figure}[tbh]
	\begin{center}
	\includegraphics[width=.5\textwidth]{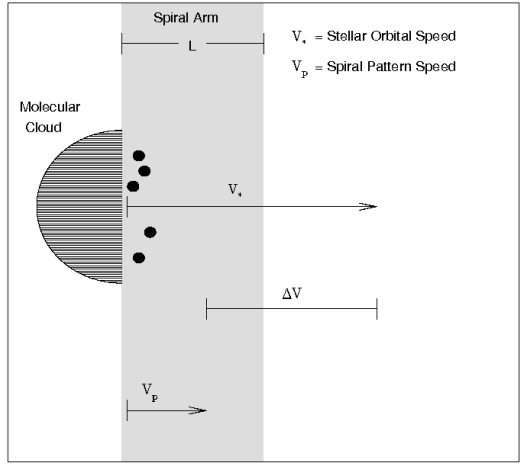}
	\end{center}
	\caption{As a molecular cloud orbits into a spiral arm, gravitational perturbations induce star formation. Since the orbital speeds of the stars are greater than the spiral arm pattern speed, the stars can, depending on their lifespans, exit the arm's leading edge. The width of the spiral arm, L, and the difference in orbital speeds, $\Delta$V, define a timescale for determining a given star's location at the end of its life.}
	\label{fig 1}
	\end{figure}

 	\begin{figure}[tbh]
	\begin{center}
	\includegraphics[width=.8\textwidth]{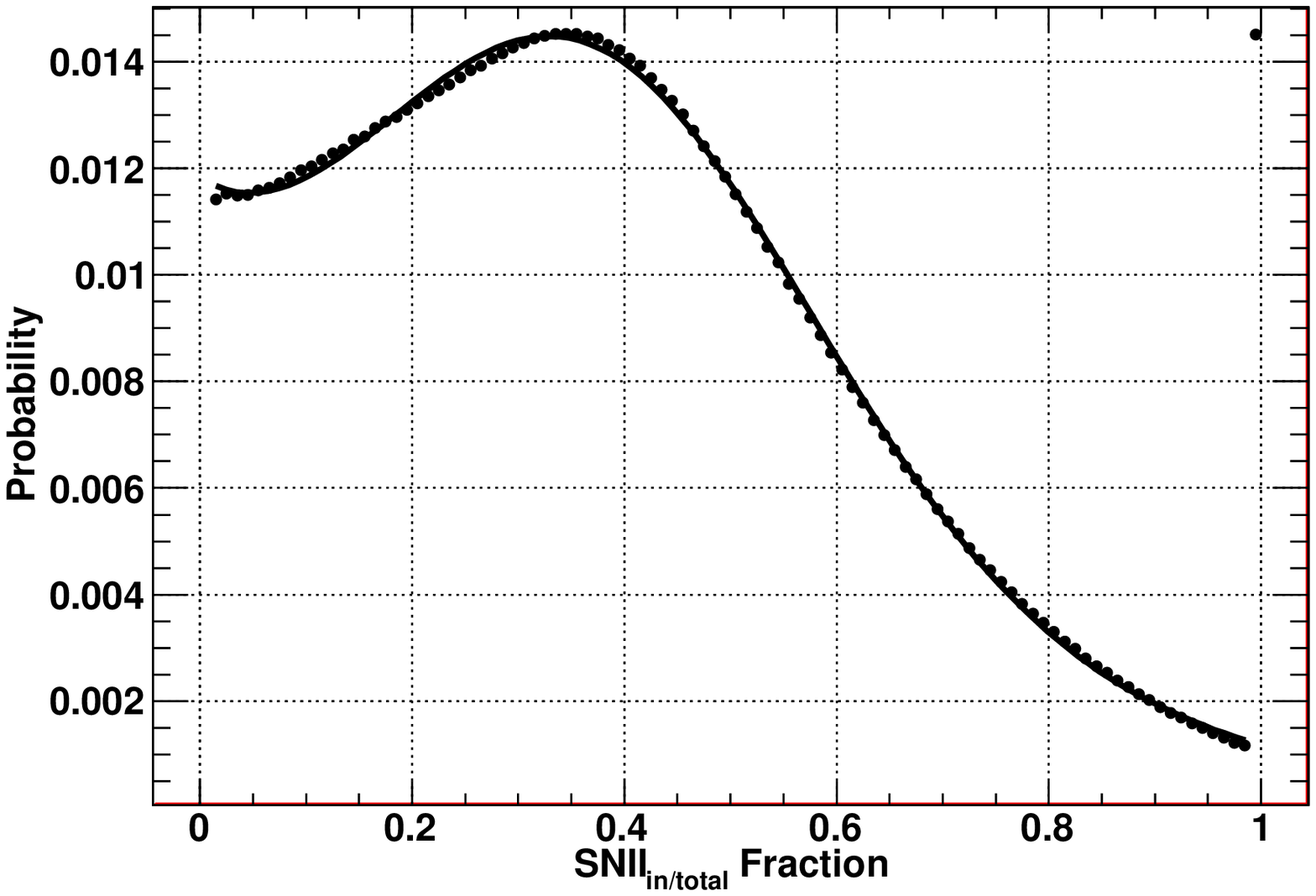}
	\end{center}
	\caption{$SNII_{in/total}$ fraction vs. Probability. The fit used to produce the PDF is plotted over the points. Discontinuities occur at values of 0.0 and 1.0 (see Results section). The probability for 0.0 is 0.0904 }
	\label{fig 2}
	\end{figure}

The PDF of $SNII_{in/total}$ is shown in Fig. 2. This distribution can be thought of as a measure of how far a given star travels from its birthplace in the spiral arm (in the arm's reference frame). Framing this distance in terms of the $SNII_{in/total}$ fraction folds the tails of the distribution into the final points at either end.

A fit of the results, normalized for use as a PDF, is given in Eqn. 1. The distribution is fit with delta functions of finite area at the endpoints and between as an exponentiated polynomial. Table 2 contains values for the coefficients.

\begin{table}[htbp]
   \centering

   \begin{tabular}{@{} lcr @{}}
      \toprule
      \multicolumn{2}{c}{Fit Parameters} \\
      \cmidrule(){1-2}
      Parameter & Value \\
      \midrule
       $\alpha_{0}$ & 1.167706502\\
       $\alpha_{1}$ & -1.23164\\
       $\alpha_{2}$ & 16.0944\\
       $\alpha_{3}$ & -42.1807 \\
       $\alpha_{4}$ & 34.2674  \\
       $\alpha_{5}$ & -9.25471   \\ 
      \bottomrule
   \end{tabular}
   \caption{}
   \label{tab:booktabs}
\end{table}

\begin{equation}
f(x)= 
\begin{cases} \text{for x=0;} & \text{$\delta(x)$}
\\
\text{for $0<x<1$;} &  \text{$\alpha_{0}e^{\sum_{i=1}^5\alpha_{i}x^i}$}
\\
\text{for x=1;} & \text{$\delta(x-1)$}
\end{cases};
Where
\int f(x)= 
\begin{cases} \text{for x=0;} &\text{0.0903762}
\\
\text{for $0<x<1$;} &  \text{0.8951162}
\\
\text{for x=1;} & \text{0.0145076}
\end{cases}
\end{equation}

\section{Discussion}

Models utilizing a concentration of SNII in the spiral arms to account for the PAMELA anomaly can be roughly tested by the integration of Eqn. 1 for the range of concentration values which allow the model to fit observed data. This provides a simple constraint on the likelihood of the concentration required by the model.

Piran et al. \cite{Piran_2009}  assume a value of 4 times more SNII in spiral arms than out (4:1). This corresponds to a $SNII_{in/total}$ fraction as used in this paper of 0.8.  The robustness of their model was not discussed with respect to this assumption, but if we assume a range of 3:1 to 5:1 still reproduces observations within their model then we can assess the likelihood.

A concentration between 3:1 and 5:1 ($\frac{3}{4} \le SN_{in/total} \le \frac{5}{6}$) corresponds to an area under the PDF of $\sim0.029$. This shows that a ratio of 4 times more SNII in a spiral arm than out is, in this model, not a robust assumption.

Further, this illustrates a general prediction of this SNII model. Namely, for a significant fraction of the parameter space, there will, in fact, be a deficiency of SNII in spiral arms; most SNII will occur in inter-arm regions.  The median calculated value is $\sim0.329$, corresponding to roughly 2 SNII in the inter-arm region for every 1 SNII inside a spiral arm.

The inputs used in this model are well constrained observationally, with the exception of spiral arm size. The physical extent of spiral arms is inherently variable; further, the edges are not sharply delineated. Coupled with possible eccentricities in the orbits of stars and molecular clouds, the range of values for the effective size of the spiral arm used in the calculations represents the only significant source of systematic error. Clearly, in the model smaller arm sizes directly bias results toward lower $SNII_{in/total}$ fractions with larger arm sizes tending in the opposite direction.

In this paper, the large range in spiral arm size (see Table 1) was used to account for these issues. However, with limits on the overdensity in the ISM required by the models of Piran\cite{Piran_2009} and Shaviv\cite{Shaviv_2009}, a better definition of spiral arm size as applied to calculating $SNII_{in/total}$ could be constructed. At that point, variances in orbital eccentricities at the Solar Galactocentric Radius, though relatively small, could be more readily accounted for to construct a more locally applicable version of this general model.

With the majority of SNII occurring outside of the progentior star's parent spiral arm, models attempting to fit the PAMELA anomaly with spatial concentrations of SNII (rather than known locations of individual SNII remnants) must take note of the level of concentration needed and it's  likelihood.

\bibliographystyle{plain}
\bibliography{Bibliography}

\end{document}